\documentclass[journal,10pt]{IEEEtran}
\usepackage{amssymb}
\usepackage{amsfonts}
\usepackage{amsmath}
\usepackage{algorithm}
\usepackage{algorithmic}
\usepackage{multicol}
\usepackage{amsmath}
\usepackage{subfigure}
\usepackage{graphicx}
\usepackage{graphicx,subfigure,amsmath,amssymb,cite}

\usepackage[dvips]{color}
\usepackage{float}
\ifCLASSINFOpdf
\else
\fi
\begin{document}

\title{Receive Antenna Selection for Secure Pre-coding Aided Spatial Modulation}

\author{Lin Liu,~Guiyang Xia,~Jun Zou,~Weibin~Zhang,~Feng Shu,~and~Jiangzhou Wang, ~\IEEEmembership{Fellow,~IEEE}
\vspace{-0.5cm}
}

\maketitle

\begin{abstract}
  In this paper, we make an investigation of receive antenna selection (RAS) strategies in the secure pre-coding aided spatial modulation (PSM) system with the aid of artificial noise. Due to a lack of the closed-form expression for secrecy rate (SR) in secure PSM systems, it is hard to optimize the RAS. To address this issue, the cut-off rate is used as an approximation of the SR. Further, two low-complexity RAS schemes for maximizing SR, called Max-SR-L and Max-SR-H, are derived in the low and high signal-to-noise ratio (SNR) regions, respectively. Due to the fact that the former  works well in the low SNR region but becomes worse in the medium and high SNR regions while the latter also has the similar problem, a novel  RAS strategy Max-SR-A  is proposed to cover all SNR regions.
  Simulation results show that the proposed Max-SR-H and Max-SR-L schemes approach the optimal SR  performances of the exhaustive search (ES) in the high and low SNR regions, respectively. In particular,  the SR performance of the proposed Max-SR-A is close to that of the optimal ES and better than that of the random method in almost all SNR regions.
\end{abstract}

\begin{IEEEkeywords}
 Secure spatial modulation, antenna selection, pre-coding, secrecy rate, finite-alphabet inputs.
\end{IEEEkeywords}

\vspace{-0.3cm}
\section{Introduction}

Wireless communication is usually prone to eavesdropping and active malicious attacks due to its broadcast characteristics. Secure spatial modulation (SSM), as enhanced SM  \cite{SM2008Mesleh} with secure capacity, is attracting ever-increasing research interest from academic world and industry due to its high energy efficiency. It is suitable for low-power-consumption scenarios like internet of things and wireless sensor networks.

Unlike the directional modulation (DM), SSM naturally suits the multi-path fading channel and the DM can only be applied in the line-of-sight scenarios \cite{shu2016robust}.
In the recent years, the researches on SSM focus on the following several aspects: pre-coding \cite{Onthesecrecy2012Guan}, artificial noise (AN) injection \cite{SE2015Wang}, transmit antenna selection (TAS)\cite{TwoHighPerformance2018Shu}\cite{Xia2019Antenna} and power allocation \cite{shu2019high}. The authors in \cite{Onthesecrecy2012Guan} proposed a pre-coding scheme to improve secrecy rate (SR) for SM systems. By projecting the AN into the null space of the legitimate channel, a secrecy enhancement scheme was proposed in \cite{SE2015Wang}. In \cite{TwoHighPerformance2018Shu}, the authors investigated a SSM system, of which two extremely low-complexity TAS schemes were proposed. As a powerful way of enhancing the security, some power spitted schemes \cite{shu2019high} between the legitimate signal and AN were also devised.

The above literature only concentrates on the transmit techniques for SSM. However, there is very little literature concerning the secure pre-coding SM (PSM). First, let us review literature about SM without taking security into account. The concept of PSM was proposed in \cite{zhang2015errorPSM} to achieve the goal of the low-detection complexity. Considering that the number of the receive antennas (RAs) is not a power of two, the authors in \cite{EfficientReceive2018Wen} proposed two efficient RA selection (RAS) schemes for the purpose of improving the bit error rate performance in PSM networks. Moreover, the authors in \cite{TPAided2016Wu} investigated the physical layer security performance of PSM schemes via pre-coding optimization by jointly exploiting the power difference between the desired user and eavesdropper.

To the best of our knowledge, there is no research of investigating how to improve the security of PSM by using RAS. Our main contributions are summarized as follows:
\begin{enumerate}
 \item  To reduce the computational complexity of optimizing the RAS, two low-complexity RAS strategies, called the maximizing SR in the low SNR (Max-SR-L) and the maximizing SR in the high SNR (Max-SR-H), are proposed. The former and the latter can achieve a perfect SR performances in the low and high SNR regions, respectively. In accordance with simulation results, the proposed two methods can outperform the random method in terms of the SR performance. More importantly, their extremely low-complexities are very of importance for practical applications.
 \item  To address the problem that there exist a large SR performance loss for the proposed Max-SR-L and Max-SR-H in the medium SNR region, we  propose a novel method maximizing SR at the all SNR region (Max-SR-A), which can perform well for all the SNR regions. Simulation results show that, compared to Max-SR-L and Max-SR-H, the proposed Max-SR-A performs better in the medium SNR region in terms of SR. In addition, it can achieve an acceptable SR performance over the whole range of SNR, but at the cost of a slightly higher computational complexity than the previous two methods.
\end{enumerate}

\emph{Notations:} Matrices, vectors, and scalars are denoted by letters of bold uppercase, bold lowercase, and lowercase, respectively. $\mathbb{C}^{M \times N}$ indicates a complex matrix with $\!M \times N\!$ dimensions. Signs $(\cdot)^{-1}$,~$(\cdot)^{H}$ and $\|\cdot\|_{F}$ denote matrix inverse, conjugate transpose and Frobenius norm. $\mathbb{E}\{\cdot\}$ stands for the expectation  operation. The operators $\lfloor\cdot\rfloor$ and $\|\!\cdot\!\|$ indicate the floor function and euclidean norm. $C(n,k)$ is the binominal coefficient. $\mathcal{CN}\left(\mathbf{\mu},\mathbf{A}\right)$ indicates a complex Gaussian distribution with mean $\mathbf{\mu}$ and covariance matrix $\mathbf{A}$.

\vspace{-0.3cm}
\section{System Model}
\begin{figure}[h]
\vspace{-3pt}   
\setlength{\abovecaptionskip}{-3pt}   
\setlength{\belowcaptionskip}{-3pt}   
\centering
\includegraphics[width=0.48\textwidth]{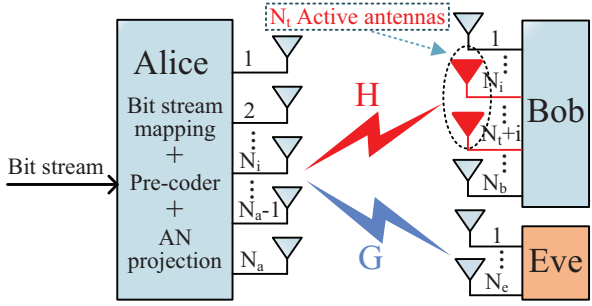}\\
\caption{System model of  secure PSM with RAS scheme.}\label{system-modelfig}
\end{figure}

Fig. \ref{system-modelfig} sketches a secure PSM system with RAS. Here, there is a transmitter (Alice) with $N_a$ transmit antennas, a legitimate receiver (Bob) with $N_b$ RAs, and an eavesdropper (Eve) with $N_e$ RAs. It is assumed that $N_b$ is not a power of two, thus we have to select $ N_{t} = 2^{\lfloor log_{2}^{N_b}\rfloor }$ out of $N_b$ RAs for mapping the bits to antenna index. Notice that there are $K=C(N_{b}, N_{t})$ possible patterns in total, represented as $\left\{\Omega_{1}, \cdots, \Omega_{K}\right\}$, where the $\Omega_{k}$ denotes the antenna set of the $k$th pattern.

After Bob chooses one pattern from the pattern set $K$, the transmitted symbol $\mathbf{x}$ with the aid of AN is given by
\begin{equation}
\mathbf{x} =\sqrt{\rho_1 P_S} \mathbf{P}_k \mathbf{e}_n s_m + \sqrt{\rho_2 P_S}\mathbf{P}_{AN}\mathbf{z},
\end{equation}
where $\mathbf{e}_n$ is the $n$th column of $\mathbf{I}_{N_b}$ for $ n \!\in\! \{1,2,\cdots,N_{t}\}$, indicating that the $n$th RA is activated at Bob; $s_m$ denotes the input symbol from an $M$-ary APM constellation. Besides, $\mathbf{z}$ is the random AN vector, $P_S$ is the total power constraint,  $\rho_1$ and $\rho_2$ are the power allocation factors with $\rho_1\!+\!\rho_2 \!=\! 1$. Assuming that the channel status information of Bob is available at Alice, the useful signal's pre-coding matrix $\mathbf{P}_k$ and the AN project matrix $\mathbf{P}_{AN}$ can be implemented by Alice.

Let $\mathbf{H} \in \mathbb{C}^{{N_{b}\times N_{a}}}$ and $\mathbf{G}\in \mathbb{C}^{N_{e} \times N_{a}}$ be the channel matrices from Alice to Bob and Alice to Eve, respectively. Each element of $\mathbf{H}$ and $\mathbf{G} $ obeys the circularly symmetric complex-valued Gaussian distribution. Accordingly, the received signals at Bob and Eve can be expressed by
\begin{equation}
\mathbf{y}_b \!=\!\mathbf{T}_{k}\mathbf{H} \mathbf{x} \!+\! \mathbf{n}_b\!=\! \sqrt{\rho_1 P_S} \beta_k \mathbf{e}_n b_m \!+\! \mathbf{n}_b,
\end{equation}
\begin{equation}
\mathbf{y}_e \!=\!\mathbf{G}\mathbf{x}\!+\!\mathbf{n}_e\!=\! \sqrt{\rho_1 P_S}\mathbf{G} \mathbf{P}_{k} \mathbf{e}_n b_m\!+\! \sqrt{\rho_2 P_S}\mathbf{G} \mathbf{P}_{AN} \mathbf{z} \!+\! \mathbf{n}_e,
\end{equation}
where $\mathbf{T}_{k}\!\in\!\mathbb{R}^{N_{t}\!\times\!N_{b}}$ is RAS matrix, constituted by the specifically selected $N_t$ rows of $I_{N_{b}}$. Assume that $\mathbf{H}_{k}\!=\!\mathbf{T}_{k}\mathbf{H} $. $\mathbf{P}_k\!=\!\beta_k \mathbf{H}_{k}^{H}\left(\mathbf{H}_{k}\mathbf{H}_{k}^{H}\right)^{-1} $, where $\beta _k\!=\!\sqrt{1/tr((\mathbf{H}_{k}\mathbf{H}_{k}^{H})^{\!-\!1})}$. $\mathbf{P}_{AN}\!=\!\frac{1}{\mu}(\mathbf{I}_{N_a}\!-\!\mathbf{H}_{k}^{H} (\mathbf{H}_{k}\mathbf{H}_{k}^{H} )^{-1} \mathbf{H}_{k})$, where $\mu\!=\!\left\|\mathbf{I}_{N_a}\!-\!\mathbf{H}_{k}^{H}(\mathbf{H}_{k}\mathbf{H}_{k}^{H})^{-1}\mathbf{H}_{k}^{H}\right\|_{\mathrm{F}}$. It is obvious that $\mathbf{T}_k\mathbf{H} \mathbf{P}_{AN} \!=\! \mathbf{0}$, so that the AN has no effect on Bob. Additionally, $\mathbf{n}_b \!\sim \!\mathcal{CN} (0, \sigma_{b}^{2} \mathbf{I}_{N_{t}})$ and $\mathbf{n}_e \!\sim\! \mathcal{CN}(0, \sigma_{e}^{2} \mathbf{I}_{N_{e}})$ are additive white gaussian noise.
As such, the average SR is defined as
\begin{equation} \label{MonteCarlorateRS}
{R}_{s}=\mathbb{E}_{\mathbf{H}, \mathbf{G}}\left(\left[I\left(\mathbf{x};\mathbf{y}_{b}|\mathbf{H},\mathbf{T}_{k},\mathbf{P}\right)-I\left(\mathbf{x};\mathbf{y}_{e}|\mathbf{G},\mathbf{P} \right)\right]^{+}\right),
\end{equation}
where $[a]^{+}\!=\!\max\{a,0\}$. $I\left(\mathbf{x};\mathbf{y}_{b} | \mathbf{H},\mathbf{T}_{k},\mathbf{P}\right)$ and $I\left(\mathbf{x}; \mathbf{y}_{e} | \mathbf{G} ,\mathbf{P} \right)$ represent the average mutual information of the Alice-to-Bob and Alice-to-Eve channels, respectively.

Considering the discrete-input continuous-output memoryless channel, the mutual information for Bob is expressed as
\begin{equation} \label{mutual_information_Bob}
\begin{split} I\left(\mathbf{x} ; \mathbf{y}_{b} | \mathbf{H},\mathbf{T}_{k},\mathbf{P}\right) = \log_{2}{MN_t} - \frac{1}{M N_t} \sum_{m=1}^M \sum_{n=1}^{N_t} \mathbb{E}_{\mathbf{n}_b} \\ \left\{ \log_2 \sum_{m'=1}^M \sum_{n'=1}^{N_t}  \exp \left(\frac{\|\mathbf{n}_b\|^2 - \|\mathbf{\delta}_{m, n}^{m', n'} + \mathbf{n}_b \|^2}{\sigma_b^2} \right)\right\},
\end{split}
\end{equation}
where $\delta_{m, n}^{m', n'} = \sqrt{\rho_{1}P_{S}}\beta_k(\mathbf{e}_n s_m -\mathbf{e}_{n'} s_{m'})$.

Due to the effect of the injected AN, the received noise at Eve is a colored noise. Then, a whitening filter $\mathbf{W}^{-1/2}$ is required and given by $ \mathbf{W} \!=\! \rho_{2}P_{S} \mathbf{G}\mathbf{P}_{AN} \mathbf{P}_{AN}^{H} \mathbf{G}^{H}\!+\!\sigma_{e}^{2} \mathbf{I}_{N_{e}}$. Therefore, the mutual information for Eve is
\begin{equation}
\begin{split} I(\mathbf{x}; \mathbf{y}_{e} | \mathbf{G} ,\mathbf{P}) = \log_{2}{MN_t} - \frac{1}{M N_t} \sum_{m=1}^M \sum_{n=1}^{N_t} \mathbb{E}_{\mathbf{n}_e^{'}} \\ \quad \left\{ \log_2 \sum_{m'=1}^M \sum_{n'=1}^{N_t}  \exp(\|\mathbf{n}_e^{'}\|^2 - \|\alpha_{m, n}^{m', n'} + \mathbf{n}_e^{'} \|^2 )\right\},
\end{split}
\end{equation}
where $\alpha_{m, n}^{m', n'} \!=\! \sqrt{\rho_{1}P_{S}} \mathbf{W}^{\frac{-1}{2}}\mathbf{GP}_k(\mathbf{e}_n s_m \!-\! \mathbf{e}_{n'} s_{m'})$ and $\mathbf{n}_e^{'} \!=\! \mathbf{W}^{-\frac{1}{2}}(\sqrt{\rho_2 P_S} \mathbf{GP}_{AN}\mathbf{z} + \mathbf{n}_e) $. Thus, $\mathbf{n}_e^{'} \!\sim\! \mathcal{CN}\left(0, \mathbf{I}_{N_{e}}\right)$. Herein, we assume that $ \mathbf{G}$ is also available at Alice \cite{Principles2014Mukherjee}, then the SR maximization problem over TAS is cast as
\begin{equation} \label{ESRSMon}
\begin{array}{cl}
{\text{~max~~~~~~~}} &{R_{s}} \\ {\text{subject~to}} &{\mathbf{T}_{k} \in\left\{\mathbf{T}_{1}, \mathbf{T}_{2}, \ldots, \mathbf{T}_{K}\right\}}.
\end{array}
\end{equation}
Note that evaluating  the SR in (\ref{ESRSMon}) requires an extremely large computational amount and is computationally prohibitive due to the use of the monte-carlo method.

\vspace{-0.3cm}
\section{Proposed RAS strategies}
In this section,  we first aim to reduce the computational complexity of evaluating $R_{s}$ in (\ref{ESRSMon}). To further mitigate the computational burden, three low-complexity RAS schemes, namely Max-SR-L, Max-SR-H and Max-SR-A, are proposed.

To reduce the computational complexity of the exact SR value, a closed-form approximation to the mutual information can be employed, namely the cut-off rate \cite{Xia2019Antenna}\cite{Aghdam2017Joint}, is given by
\begin{equation} \label{cut-of-rate-Bob}
I_{b}^0  \!=\! -\log_2\frac{1}{(MN_t)^2}\sum_{m'=1}^M\sum_{n'=1}^{N_t}\sum_{m=1}^M\sum_{n=1}^{N_t} \exp \left(\frac{ -\|\delta_{m, n}^{m', n'}\|^2}{4\sigma_b^2} \right)
\end{equation}
\begin{equation} \label{approximateRateEve}
I_{e}^0  = - \log_2\frac{1}{(MN_t)^2}\sum_{m'=1}^M\sum_{n'=1}^{N_t}\sum_{m=1}^M\sum_{n=1}^{N_t} \exp \left(\frac{ -\|\alpha_{m, n}^{m', n'}\|^2}{4} \right),
\end{equation}
with the approximate SR  as $R_{s}^{\prime}=I_{b}^0 - I_{e}^0$. Upon replacing $R_s$ by $R_{s}^{\prime}$, it will save a large computational amount.  Although $R_{s}^{\prime}$ is not the achievable SR, its efficiency has been repeatedly demonstrated in \cite{Xia2019Antenna}.
Then, the optimization problem in \eqref{ESRSMon} is reduced to
\begin{equation} \label{argmaxASR}
\mathbf{T}_{k^{*}}=\underset{\mathbf{T}_{k} \in\left\{\mathbf{T}_{1}, \mathbf{T}_{2}, \ldots, \mathbf{T}_{K}\right\}}{\arg \max }~~ R_{s}^{\prime},
\end{equation}
where $R_{s}^{\prime}$ has a closed form. However, \eqref{argmaxASR} is still an intractable optimization problem due to its prohibitive computational complexity. Below, we propose some methods to further mitigate this computational burden.

\vspace{-0.3cm}
\subsection{Proposed Max-SR-L RAS Method}

Firstly, the nature of $R_{s}^{\prime}$ will be investigated as the SNR tends to infinitely small, i,e., $\sigma^2_e \rightarrow \infty$. It can be noted that the value of expression $\|\mathbf{\alpha}_{m, n}^{m', n'}\|^2 $ is closed to $0$, with the aid of the first-order Taylor expansion, we have $\exp (\frac{ -\|\mathbf{\alpha}_{m, n}^{m', n'}\|^2}{4} ) \!\approx\! 1 \!-\! \frac{\| \mathbf{\alpha}_{m, n}^{m', n'}\|^2}{4} $ and $\log_2(1\!-\!x) \!\approx\! -x/\ln 2$. As such, an approximation to Eve's mutual information in the low SNR can be expressed as
\begin{equation} \label{LOW_SNReqe}
\begin{aligned}
& I_{e}^l  \!\approx\!-\log_2 \left(1\!-\!\frac {1}{ 4(MN_{t})^2 } \sum_{m'=1}^M\sum_{n'=1}^{N_t}\sum_{m=1}^M\sum_{n=1}^{N_t}{\|\mathbf{\alpha}_{m, n}^{m', n'}\|^2}\right)\\
 & \!\leq\! \frac {\rho_1 P_s\|\mathbf{Q}\|_{2}^{2}}{ 4(MN_{t})^2 \ln 2 }\sum_{m'=1}^M\sum_{n'=1}^{N_t}\sum_{m=1}^M\sum_{n=1}^{N_t}{\|\mathbf{e}_{n}s_{m}\!-\!\mathbf{e}_{n'} s_{m'}\|^2},
\end{aligned}
\end{equation}
where $ \|\mathbf{\alpha}_{m, n}^{m', n'}\|^2 \!\leq\! \rho_{1}P_{S}\|\mathbf{Q}\|_{2}^{2}\|\mathbf{e}_{n}s_{m}\!-\!\mathbf{e}_{n'} s_{m'}\|^2 $ with $\mathbf{Q} = \mathbf{W}^{-\frac{1}{2}}\mathbf{GP}_k $. Similarly, the mutual information term $I_b$ can be approximated as
\begin{equation} \label{LOW_SNReqb}
I_{b}^l \approx \frac {\rho_{1}P_{S} \beta_k^2}{4\sigma_b^2(MN_{t})^2 \ln 2 } \sum_{m'=1}^M\sum_{n'=1}^{N_t}\sum_{m=1}^M\sum_{n=1}^{N_t}{ \|\mathbf{e}_{n}s_{m}\!-\!\mathbf{e}_{n'} s_{m'}\|^2 }.
\end{equation}
Upon using the difference of (\ref{LOW_SNReqe}) and (\ref{LOW_SNReqb}), an approximation to the SR is
\begin{equation} \label{lowSNRrs1}
  R_{s}^l =I_{b}^l - I_{e}^l.
\end{equation}
Replacing the objective function in (\ref{argmaxASR}) by (\ref{lowSNRrs1})  yields
\begin{equation} \label{lowSNRrs}
\mathbf{T}_{k^{*}}=\underset{\mathbf{T}_{k} \in\left\{\mathbf{T}_{1}, \mathbf{T}_{2}, \ldots, \mathbf{T}_{K}\right\}}{\arg \max }~~ {\beta_k^2}{\sigma_b^{\!-\!2}} - \|\mathbf{Q}\|_{2}^{2}.
\end{equation}
It is apparent that the objective function in \eqref{lowSNRrs} has a much simpler form than that in \eqref{argmaxASR}. Actually, the two-layer summation over all possible transmit symbols has been removed from the derived objective function in \eqref{lowSNRrs}. Therefore, the computational complexity is significantly reduced.

\vspace{-0.3cm}
\subsection{Proposed Max-SR-H RAS Method}

It should be pointed out that the proposed scheme in \eqref{lowSNRrs} has a low complexity, but it merely suits for the low-SNR region. As we will show in the next section, its SR performance will be seriously deteriorated in the high SNR region. Here, we propose a new scheme suiting for the high-SNR region.  When $\sigma_b^2\rightarrow0$, i.e., in the high SNR region, we have
\begin{equation} \nonumber
\mathbb{E}_{\mathbf{n}_b} \left\{ \log_2 \sum_{m'=1}^M \sum_{n'=1}^{N_t}  \exp \left(\frac{\|\mathbf{n}_b\|^2 \!-\! \|\mathbf{\delta}_{m, n}^{m', n'} \!+\! \mathbf{n}_b \|^2}{\sigma_b^2} \right)\right\} \rightarrow 0 .
\end{equation}
Thus, the achievable rate obtained by Bob is close to $\log_2MN_t$, which can be observed from (\ref{mutual_information_Bob}). From this perspective, choosing different RA combinations will result in little effect on the achievable rate of Bob. Considering that the SR is calculated from the difference between the achievable rates of Bob and Eve. Because Bob's rate is almost unchanged,  the SR can be increased by reducing the achievable rate of Eve. From the perspective of information theory, reducing Eve's achievable rate can be achieved by minimizing its received SNR of the useful signals at Eve. Consider that the AN and thermal noise are whitened after the received signal passes by taking the benefit of the whitening filter $\mathbf{W}^{\frac{-1}{2}}$. The corresponding optimization problem can be described as
\begin{equation} \label{MaxSR_H}
\begin{array}{cl}
{\min}~~~~~~~ &\|\mathbf{W}^{\frac{-1}{2}}\mathbf{GP}_k\|_F^2 \\ {\text{subject to}}~ & {\mathbf{T}_{k} \in\left\{\mathbf{T}_{1}, \mathbf{T}_{2}, \ldots, \mathbf{T}_{k}\right\}}.
\end{array}
\end{equation}
As such, a large number of the computational load will be saved by taking the benefit of the simplified objective function of \eqref{MaxSR_H}. Actually, the Max-SR-H scheme may coordinate with the Max-SR-L together to avoid the disadvantage of the SR performance degradation in the high SNR region.

\vspace{-0.3cm}
\subsection{Proposed Max-SR-A RAS Method}

In the previous two subsections, two simple approximations to the SR expression considered only the low and high SNR regions, respectively. How about the medium SNR region? In this subsection, a new  approximation is presented to be independent of SNR regions. Then, a new algorithm, called Max-SR-A, is proposed to conquer this disadvantage.

Assume that the set of all possible values of $\|\mathbf{e}_n s_m \!-\! \mathbf{e}_{n'} s_{m'}\|^2$ is replaced by $\{ d_1, d_2,\cdots, d_J \}$ and $\{ f_1,f_2,\cdots, f_J \}$ where the former denotes the values and the latter denotes its probability mass function. For the determined transmission mode, $d_i$ and $f_j$ are calculated in advance just once. The cut-off rate in (\ref{cut-of-rate-Bob}) for Bob can be simplified as
\begin{equation} \label{Ibrateh}
 I_{b}^a = -\log_2\sum_{j=1}^J f_j \exp \left(\frac{-\rho_1 P_s \beta_k^2}{4 \sigma_b^2} d_j \right).
\end{equation}
Due to the SM nature, the number of $d_j$ can be greatly reduced because of only a unique '1' component. Thus the computational complexity of the cut-off rate can be decreased. According to the definition of the matrix norm induced by vector norms, we get $\|\mathbf{A}x\|^{2} \!\leq\!  \|\mathbf{A}\|_{2}^{2}\|x\|^{2}$, and the inequality $\exp (\frac{-\|\alpha_{m, n}^{m', n'}\|^2} {4} ) \geq \exp (\frac{-\rho_1 P_s}{4}\| \mathbf{Q}_k\|_2^2 \|\mathbf{e}_{n}s_{m}\!-\!\mathbf{e}_{n'} s_{m'}\|^2)$ can be derived.
Thus, the rate $I_e^h$ is given by
\begin{equation} \label{Ierateh}
I_{e}^a \approx -\log_2\sum_{j=1}^J f_j \exp \left(\frac{-\rho_1 P_s}{4}\|\mathbf{Q}\|_{2}^2 d_j \right),
\end{equation}
Taking the difference between \eqref{Ibrateh} and \eqref{Ierateh} as the new alternative, the SR approximation can be further reduced to
\begin{equation}
R_{s}^a =\log_2{\frac{\sum_{j=1}^J f_j \exp \left(\frac{-\rho_1 P_s}{4}\|\mathbf{Q}\|_{2}^2 d_j \right)} {\sum_{j=1}^J f_j \exp \left(\frac{-\rho_1 P_s \beta_k^2}{4 \sigma_b^2} d_j \right)} },
\end{equation}
which gives the following optimization problem
\begin{equation} \label{maxSR_A}
\mathbf{T}_{k^{*}}=\underset{\mathbf{T}_{k} \in\{\mathbf{T}_{1}, \mathbf{T}_{2}, \ldots, \mathbf{T}_{K}\}}{\arg \max }~~ R_{s}^{a}.
\end{equation}
Clearly, the computational complexity of  \eqref{maxSR_A} is much higher than those of  Max-SR-L and Max-SR-H. This is mainly owing to the fact that calculating the objective function $R_{s}^{a}$ requires larger computational load than those of  Max-SR-L and Max-SR-H. However, this method is capable of harvesting stable SR gains over  Max-SR-L and Max-SR-H for the medium SNR region, which will be shown in the simulation section.

\vspace{-0.3cm}
\subsection{Complexity Analysis and Comparison}

Now, we make a complexity comparison concerning the above methods. Notice that there are totally $K = C(N_b,N_t)$ possible RAS patterns, and the complexity of exponential and logarithm operations are omitted. According to the floating-point operations (FLOPs) in matrix-vector calculus \cite{hunger2005floating}, the original Max-SR method using Monte-Carlo simulation costs about $C_{SR}\!=\!KM^{2} N_{t}^{2}N_{samp}(2N_eN_t \!+\!6N_t \!+\! 5N_e \!-\! 1) $ FLOPs, where $N_{samp}$ is the number of noise samples for evaluating the SR. The proposed Max-SR-L doesn't require exhaustive search over all  symbols, and its complexity is reduced to $ C_{1L}\!=\!K(2N_e^2 N_a \!+\! 2N_e N_a N_t \!+\! 2N_e^2N_t \!-\! N_eN_a \!-\! N_eN_t \!+\! \frac{1}{3}N_e^3 \!-\! \frac{4}{3}N_e + 3) $. Similarly, the computation complexity of Max-SR-H is $ C_{2H} \!=\! K (2N_e^2 N_a \!+\! 2N_e N_a N_t \!-\! N_eN_a \!+\! 2N_eN_t \!-\! 1) $ FLOPs. The complexity of Max-SR-A is $C_{3A}\!=\!K (2N_e^2 N_a \!+\! 2N_e N_a N_t \!+\! 2N_e^2N_t \!-\! N_eN_a \!-\! N_eN_t \!+\! \frac{1}{3}N_e^3 \!-\! \frac{4}{3}N_e \!+\! J*6 \!+\! 7)$ FLOPs, where the value of $J$ is only related to the specific modulation constellation. Generally, $N_{samp}$ is always a large number and $N_a \!>\! N_t$, $N_t \!=\! N_e$.
Therefore, we have the complexity order as: Max-SR $>$ Max-SR-A $>$ Max-SR-L $>$ Max-SR-H.

\vspace{-0.3cm}
\section{Numerical Results and Discussions}
In this section, we present simulation results to  evaluate the performance of the above schemes. For a fair comparison, it is assumed that  $\sigma_b^2=\sigma_e^2$. In addition, the total transmit power is set as $P_S\! =\! 1$, and quadrature phase shift keying (QPSK) modulation is used. The metric for quantifying the secrecy performance is the ergodic SR obtained by averaging over 2000 random channel realizations.
\begin{figure}[h]
\vspace{-9pt}  
\setlength{\abovecaptionskip}{-3pt}   
\setlength{\belowcaptionskip}{-9pt}   
\centering
\includegraphics[width=0.46\textwidth]{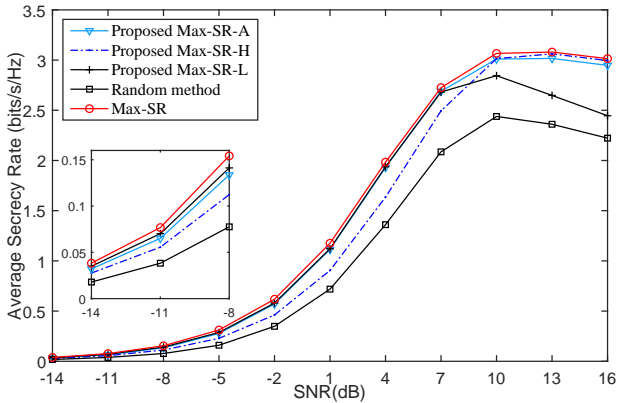}\\ 
\caption{Average SR versus SNR ($N_a=10,N_b=7,N_t=N_e=4,\rho_1=0.5$ ).}\label{fig-security-rate}
\end{figure}
\vspace{-0.2cm}

Fig.~\ref{fig-security-rate} plots the average SR performance versus SNR for the proposed three methods, where the random selection method is invoked as the benchmark. It is seen from Fig.~\ref{fig-security-rate} that the proposed three methods  have significant performance improvements over the random one. Particularly, the proposed Max-SR-L  is close to the optimal Max-SR in the low SNR region. However, its performance degrades severely in the high SNR region. By contrast, the SR performance of the proposed Max-SR-H  is close to that of the Max-SR scheme in the high SNR region, while its performance gradually decreases as SNR moves from medium to low.
The proposed Max-SR-A  is close to the Max-SR for all SNRs whereas it is much better than both the Max-SR-L and Max-SR-H in the medium SNR region, and slightly worse than Max-SR-L and Max-SR-H in the low and high SNR regions, respectively.
\begin{figure}[h]
\vspace{-9pt}  
\setlength{\abovecaptionskip}{-3pt}   
\setlength{\belowcaptionskip}{-9pt}   
\centering
\includegraphics[width=0.46\textwidth]{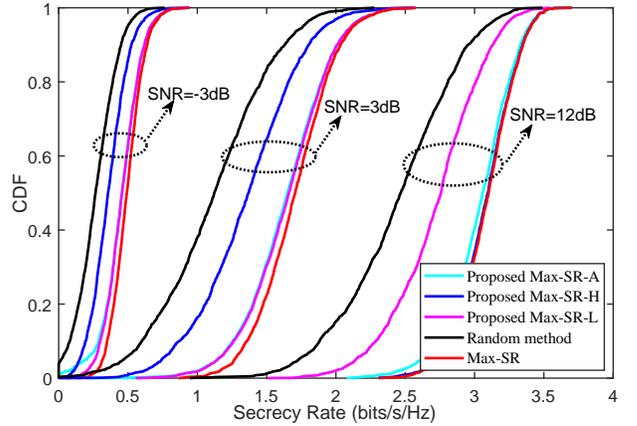}\\
\caption{CDF of SR.}\label{CDF}
\end{figure}
Fig. \ref{CDF} shows the cumulative distribution function (CDF) of SR for the proposed three RAS methods with three different values of SNR: -3dB, 3dB and 12dB. From Fig. \ref{CDF}, we find the consistent SR performance tendency as Fig.~\ref{fig-security-rate}.

\vspace{-0.4cm}
\section{Conclusion}
In this paper, we have investigated the RAS strategies of the secure PSM systems. To reduce the computational complexity and improve the SR performance, three high performance TAS strategies, called Max-SR-L, Max-SR-H and Max-SR-A, have been proposed. Simulation results showed that the proposed Max-SR-L and Max-SR-H  can attain substantial SR gains over the random method in the low and high SNR regions with extremely low-complexities, respectively. The proposed Max-SR-A  is capable of achieving the near-optimal SR performance over almost all SNR regions and  avoiding the SR performance losses of the Max-SR-L and Max-SR-H in the high and low SNR regions, respectively.

\vspace{-0.3cm}
\bibliographystyle{IEEEtran}
\bibliography{RAS}
\end{document}